\documentclass[11pt,a4paper]{article}
\usepackage[margin=2.8cm,bottom=3.5cm]{geometry}
\usepackage{amsmath}
\usepackage{amssymb}
\usepackage{color}
\pdfoutput=1 
\usepackage{graphicx}
\usepackage{float}
\usepackage{subcaption}
\usepackage[T1]{fontenc}
\usepackage[utf8]{inputenc}

\usepackage[backend=biber, sorting=none, style=numeric-comp]{biblatex}
\title{Scalar and spinor fields in gravitating cosmic string spacetimes}

\author{
        Marcos Silva \\
        Departamento de F{\'i}sica, Universidade Federal de Pernambuco,\\
        Av. Prof. Moraes Rego, 1235, Recife - PE - 50670-901, Brazil\\
        marcos.viniciussantos@ufpe.br
            \and
        Azadeh Mohammadi\\
        Departamento de F{\'i}sica, Universidade Federal de Pernambuco,\\
        Av. Prof. Moraes Rego, 1235, Recife - PE - 50670-901, Brazil\\
        azadeh.mohammadi@ufpe.br
}

\begin{document}
\maketitle

\begin{abstract}

We study the scattering behavior of scalar and spinor fields in the background of a gravitating cosmic string spacetime. The model explored here for the background vortex is non-abelian, becoming abelian in an appropriate limiting case. We adopted the formalism we developed in \cite{silva2021scattering}, modifying the standard partial wave approach. We apply the method for a scalar and also a fermion field interacting with the background spacetime with a nontrivial asymptotic structure. The spacetime metric, obtained numerically in \cite{de2015gravitating}, forms the basis of our state-of-the-art numerical study. We make an exhaustive analysis and compare all the results in the non-abelian model with the corresponding abelian one for both massless and massive fields. We analyze the field configuration's total cross-section and angular profile at small and large distances from the core.
We show that the total cross-section oscillates with the incident momentum of the wave, as anticipated in \cite{silva2021scattering}, and also, the angular profile can be explained reasonably well with a Fraunhofer diffraction pattern, especially for the scalar field scattering.

\end{abstract}

\section{Introduction}

Kibble's seminal 1971 work \cite{kibble1976topology} established that cosmological phase transitions generically produce topological defects, including monopoles, domain walls, and cosmic strings. While monopole solutions have been extensively explored in Grand Unified Theories (GUTs) \cite{t1974magnetic, van1976regular, zeldovich1978concentration, preskill1979cosmological, cho1997monopole}, cosmological constraints increasingly disfavor their prevalence \cite{milton2006theoretical, ueno2012search, barrie2021searching, acharya2022search}. In contrast, domain walls and cosmic strings remain observationally viable. Domain walls, for instance, offer compelling explanations for pulsar timing array signals \cite{jiang2022implications, ferreira2023gravitational, afzal2023nanograv} and primordial black hole formation \cite{liu2020primordial}, and may contribute to the stochastic gravitational wave background (SGWB) \cite{zhang2023nano}.
Cosmic strings, our focus, may seed intermediate-mass black holes \cite{bradenberger2021intermediate}, participate in the early structure formation \cite{jiao2023early,duplessis2013note}, and generate distinctive SGWB signatures \cite{rybak2017semianalytic, blanco2017stochastic, sousa2020full, figueroa2020irreducible, auclair2020particle, lazarides2021cosmic, auclair2023window, zhou2022gravitational,qiu2023gravitational,rybak2022emission,rybak2025stochastic,schmitz2024gravitational,
mukovnikov2024ultrahigh}, with recent evidence suggesting they could explain NANOGrav observations \cite{blanco2021comparison}. Cosmic string networks underpin these predictions \cite{correia2019extending, gonzalez2021effective,austin1993evolution, hindmarsh2017scaling}.

Observations have progressively constrained the viable parameter space for cosmic strings. Cosmic Microwave Background (CMB) studies \cite{ade2014planck, lazanu2015constraints, hergt2017searching, hindmarsh2019type} initially bounded string tensions, though wiggly-string anisotropies may be overestimated \cite{rybak2017semianalytic}. Gravitational wave searches now provide tighter limits: LIGO-Virgo \cite{abbott2021constraints}, NANOGrav \cite{buchmuller2020nanograv,kume2024revised}, and PPTA \cite{bian2022searching} data collectively restrict the string tension $\mu$, with LISA expected to probe deeper into viable parameter space \cite{auclair2020probing, boileau2022ability}.

Early works modeled the gravitational properties of straight cosmic strings in the wire approximation, i.e., treating strings as having no width, a reasonable cosmological simplification \cite{vilenkin1981gravitational, gott1985gravitational, linet1987vortex, frolov1989gravitational} but one that introduces a singularity at the string core. This idealization describes the string as a "crack" in spacetime, where removing an angular wedge $\delta = 8\pi \mu$ creates locally flat spacetime with conical topology. Crucially, matter content plays no role in this model; only the constant energy density, $\mu$, matters.

The scenario differs for gravitating strings (or extended vortices), where internal structure influences the deficit angle. Christensen, Larsen, and Verbin \cite{christensen1999complete} provided a complete classification of static cylindrically symmetric solutions in the abelian-Higgs model, identifying regions of parameter space yielding asymptotically flat conical spacetimes (i.e., cosmic strings). Brihaye and Lubo numerically constructed these solutions \cite{brihaye2000classical}, revealing key gravitational features, such as asymptotic conical topology. Many works have confirmed that resolving the axial singularity requires asymptotic conicality with a non-vanishing curvature \cite{garfinkle1985general, dyer1995complete, hartmann2008gravitating, van2013geometry}, which may source bounded geodesic motion \cite{hartmann2010geodesic, hackmann2010test, hackmann2010complete}.

Gravitational scattering by cosmic strings has been studied across various models and particle types \cite{perkins1991scattering, spinelly2001relativistic, katanaev1999scattering, neto2020scalar}. While field solutions in the vortex exterior are tractable with standard methods, gravitational scattering proves subtler. Deser and Jackiw \cite{deser1988classical} demonstrated that the deficit angle of an ideal string causes divergences in non-relativistic scalar scattering amplitudes, demanding a modification of the asymptotic ansatz in partial-wave approaches. In a previous work \cite{silva2021scattering}, we identified analogous challenges for relativistic scalar fields in the spacetime of extended gravitating strings and proposed further adaptations to handle non-trivial asymptotics. This work aims to answer how scalar and spinor fields scatter when interacting with a fully relativistic gravitating vortex.

The paper is organized as follows: Section \ref{model} presents the gravitating string model we are considering and the corresponding spacetime metric. In section \ref{scalar}, we study a scalar field interacting with the background spacetime by outlining how the total cross-section is computed and show the scalar field results, i.e., total cross-section and field profile in the Fresnel and Fraunhofer regimes. In section \ref{fermion}, we study the scattering of a fermionic wave in the background spacetime of a generic gravitating string and derive the formula for the total cross-section. Then, we show the fermionic total cross-section and field profile in the same background spacetime shown in section \ref{model}. For both scalar and spinor fields, we employ the method developed in \cite{silva2021scattering} to find the scattering cross-section, and compare results in both representative backgrounds. Finally, in section \ref{conclusion}, we provide a summary and concluding remarks. Throughout this paper, we set $\hbar = c = 1$.

\section{The model}
\label{model}
In \cite{de2015gravitating}, the authors studied the vortex solutions of the following non-abelian model
\begin{equation}
    \mathcal{L} = \frac{1}{2}(D_\mu\varphi^a)^2+\frac{1}{2}(D_\mu\chi^a)^2 -\frac{1}{4}F_{\mu \nu}^aF^{\mu \nu a}-V(\Phi^a,\chi^a),
    \label{vortex_lagrangian}
\end{equation}
for a $SU(2)$ gauge field, $A^{a}_{\mu}$, coupled with two scalar fields $\phi^a$, $\chi^a$, interacting via the generalized Higgs potential
\begin{equation}
    V(\Phi^a,\chi^a)=\frac{\lambda_1}{4} \left[(\Phi^a)^2-\eta_1^2\right]^2+\frac{\lambda_2}{4} \left[(\chi^a)^2-\eta_2^2\right]^2+\frac{\lambda_3}{2} \left[(\Phi^a)^2-\eta_1^2\right]\left[(\chi^a)^2-\eta_2^2\right].
    \label{vortex_potential}
\end{equation}
The constant parameters $\eta_1$, $\eta_2$ measure the vacuum energy. The positive constants $\lambda_1$,  $\lambda_2$ represent the self-coupling of the Higgs fields, and $\lambda_3$ the coupling between the two scalar fields. The requirement of magnetic flux quantization along the string axis forces the non-abelian theory to include two scalar fields, as shown in \cite{nielsen1973vortex}. Notice that the third term of the potential $V(\phi^a, \chi^a)$ is the simplest interaction term respecting the boundary condition $V(r \to \infty) = 0$.

\par
The model given by the lagrangian (\ref{vortex_lagrangian}) is a straightforward generalization of the abelian-Higgs model and has the Nielsen-Olsen vortex as a particular case. In fact, it reduces to the abelian-Higgs case if we set $\lambda_2 = \lambda_3 = 0$ and $\chi \equiv 0$. Consequently, we have the abelian and the non-abelian vortices within the same model.

In \cite{de2015gravitating}, the authors coupled the lagrangian (\ref{vortex_lagrangian}) with gravity by minimizing the action
\begin{equation}
S = \int d^4x \sqrt{-g} \left( \mathcal{L} + \frac{1}{2 \kappa} R \right) ,
\end{equation}
where $\kappa = 8 \pi G$. They solved the field equations numerically taking the following ansatz for the metric
\begin{equation}
    ds^{2}=N^{2}(\rho) dt^{2}-d\rho^{2}-L^{2}(\rho) d\varphi-N^{2}(\rho) dz^2,
    \label{metric_ansatz}
\end{equation}
which is the most general cylindrically symmetric metric that is invariant under boosts along the z-axis.\\
The coordinates and constant parameters were rescaled as
\begin{equation}
\begin{gathered}
    \quad r = \sqrt\lambda_1\eta_1 \rho, \quad L(r) = \sqrt\lambda_1\eta_1 L(\rho)
    \\
    \alpha = \frac{e^2}{\lambda_1}, \quad q = \frac{\eta_1}{\eta_2}
    \\
    \beta_i = \frac{\lambda_i}{\lambda_1},\quad \gamma = \kappa \eta_1^2 ,
\end{gathered}
\label{vortex_parameters}
\end{equation}
setting the dimensionless parameters $\alpha = 1.0$, $\gamma = 0.6$ for both abelian and non-abelian cases, and also fixing $q = 1.0$, $\beta_2 = 2.0, \beta_3 = 1.0$ in the non-abelian one. To have a notion of scales, the chosen parameters means that the unit of the radial length $r$ in natural units is around $5 \times 10^9$ m.

\begin{figure}[ht]
    \centering
    \includegraphics[width =1.0\textwidth]{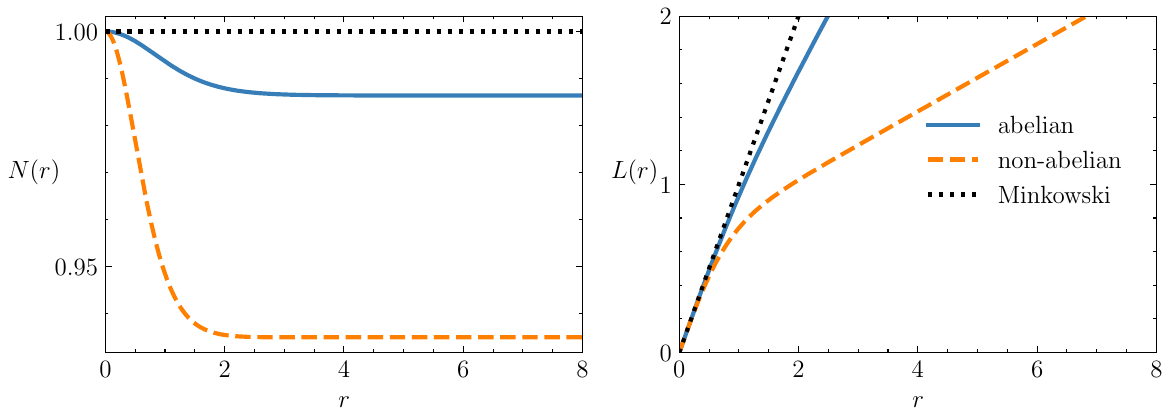}
    \caption{Metric functions $N(r)$ and $L(r)$. In both panels, solid line is the abelian case, dashed is non-abelian and dotted line is the Minkowski case.}
    \label{metric}
\end{figure}

The metric functions $N(r)$ and $L(r)$ with different parameters are reproduced in Fig. \ref{metric}.
We can observe that as $r \rightarrow 0$ the metric becomes Minkowski and as $r \rightarrow \infty$ we have a flat spacetime with
\begin{equation}
    \begin{gathered}
        N(r \rightarrow \infty) = a \\
        L(r \rightarrow \infty) = br + c,
    \end{gathered}
\label{metric_coniclimit}
\end{equation}
with spacetime constants $a, b, c$ which depend on the vortex parameters (\ref{vortex_parameters}). In the abelian case, we have approximately $a = 0.98, b = 0.64, c = 0.39$ and $a = 0.93, b = 0.20, c = 0.63$ in the non-abelian one. Hence, the deficit angle, defined by $\delta = 2 \pi (1 - b)$, is approximately $0.72\pi$ in the abelian case and $1.60 \pi$ in the non-abelian one. 

\section{Scalar field scattering}
\label{scalar}
Now, we aim to analyze the scattering of a scalar field in the gravitational potential generated by the cosmic string represented by the metric (\ref{metric_ansatz}). The first step is to solve the Klein-Gordon equation minimally coupled to gravity, given by

\begin{equation}
    (\Box + M^2)\Phi = 0,
    \label{scalar_kleingordon}
\end{equation}
where $M$ is the mass of the field, and $\Box$ is the d'Alembertian operator
\begin{equation*}
\Box = \nabla_\mu \nabla^\mu  = \frac{1}{\sqrt{-g}}\partial_\mu{(\sqrt{-g}g^{\mu\nu}\partial_\nu)},
\end{equation*}
with $g$ being the determinant of the metric.
Writing (\ref{scalar_kleingordon}) explicitly, we get
\begin{equation}
    \Bigg{\{}\frac{1}{N^2L}\Bigg{[}L\partial_{t}^2-\Big{(}2NN'L+N^2L'\Big{)}\partial_{r}-N^2L\partial_{r}^2-\frac{N^2}{L}\partial_{\varphi}^2-L\partial_{z}^2\Bigg{]} + M^2\Bigg{\}}\Phi=0,
    \label{scalar_eom1}
\end{equation}
where prime denotes derivative with respect to $r$. In order to solve \eqref{scalar_eom1} we start with the ansatz
\begin{equation}
    \Phi = e^{-iEt} e^{-ikz} \sum_{m=0}^{\infty} a_m R_{m}(r) e^{im\phi},
    \label{scalar_ansatz}
\end{equation}
where $a_m$ is a constant to be determined by the initial condition. Notice that we consider the field to have definite energy, linear and angular momenta in the $z$-direction. Replacing (\ref{scalar_ansatz}) in (\ref{scalar_eom1}), we obtain
\begin{equation}
    R_{m}''(r) + \Bigg{(}\frac{2N'(r)}{N(r)}+\frac{L'(r)}{L(r)}\Bigg{)}R_{m}'(r)+\Bigg{(}\frac{\tilde{p}^2}{N^2(r)}-\tilde M^2\Big{(}1-\frac{1}{N^2(r)}\Big{)}-\frac{m^2}{L^2(r)}\Bigg{)}R_{m}(r)=0,
    \label{scalar_eom2}
\end{equation}
where $\tilde{p}^2 = \tilde{E}^2 - \tilde{M}^2 -\tilde{k}^2$ with $\tilde{E} = E/ \sqrt{\lambda_1}\eta_1$, $\tilde{k} = k/\sqrt{\lambda_1}\eta_1$. The parameter $\tilde{p}$ can be seen as the momentum in the plane perpendicular to the string. From now on we drop the tilde for simplicity.

In the limit $r \rightarrow 0$, eq. \eqref{scalar_eom2} tends to

\begin{equation}
    R_{m}''(r) + \frac{1}{r} R_{m}'(r) + \left( p^2 - \frac{m^2}{r^2} \right)R_m(r) = 0,
    \label{scalar_eom_minkowski}
\end{equation}
which is the Bessel equation. The solution to (\ref{scalar_eom_minkowski}) is proportional to the Bessel function of the first kind, $R_m (r) = J_{|m|}(p r)$. We set $k = 0, a_m = i^{|m|}$ to have a plane wave for $\Phi$ close to the origin. So, the complete field solution at the center of the vortex is given by

\begin{equation}
\phi = e^{-iEt} \sum_{m = - \infty}^{\infty} i^{|m|} J_{|m|}(p r) e^{i m \varphi} = e^{-iEt} e^{ip r \cos\varphi}
\end{equation}
where we used the identity $J_{-n}(x) = (-1)^n J_n(x)$ for integer values of $n$ \footnote{In \cite{silva2021scattering}, the mode in Bessel function should be replaced by $|m|$ to avoid the contribution from the Neumann function.}.

On the other hand, in the limit $r \rightarrow \infty$, eq. \ (\ref{scalar_eom2}) tends to

\begin{equation}
    R^{\prime \prime}(w) + \frac{1}{w}R'(w) + \left( {p^{\prime}}^2 - \frac{(m/b)^2}{w^2} \right)R(w) = 0,
    \label{scalar_eom_infty}
\end{equation}
where $w = r + c/b$, ${p^\prime}^2 = p^2/a^2 - M^2(1 - 1/a^2) = \frac{E^2}{a^2} - \frac{k^2}{a^2} - M^2$ and prime in $R$ denotes derivative with respect to $w$.
The general solution to (\ref{scalar_eom_infty}) is $R_m = b_m J_{m'}(p^{\prime}w) + c_m Y_{m'}(p^{\prime}w)$, where $Y_{m'}$ is the Bessel function of the second kind, or Neumann function, of order $m' = m/b$. Asymptotically, this solution approaches 
\begin{align}
    R_m(r \rightarrow \infty) = b_m \sqrt{\frac{2}{\pi p^{\prime} r}} \cos\left [p^\prime w - \frac{\pi}{2}\left(m^\prime + \frac{1}{2} \right) \right] + c_m \sqrt{\frac{2}{\pi p' r}} \sin\left [p^\prime w - \frac{\pi}{2}\left(m^\prime + \frac{1}{2} \right) \right].
\end{align}
Following the common procedure, we eliminate the Neumann function, $Y_m$, by adding a phase to the Bessel function. Formally, one can set $b_m = C_m \cos(d_m)$, $c_m = C_m \sin(d_m)$ and then the asymptotic solution becomes
\begin{equation}
    R_m(r \rightarrow \infty) = C_m \sqrt{\frac{2}{\pi p^{\prime} r}} \cos \left[p^{\prime} w - \frac{\pi}{2}\left(m^\prime + \frac{1}{2} \right) + d_m(p)\right],
\label{scalar_asymp_solution}
\end{equation}
where $C_m(p)$ and $d_m(p)$ are determined by how the spacetime approaches the asymptotic limit \eqref{metric_coniclimit}; in pratice they are determined numerically. We can now determine the scattering amplitude, from which we derive the total cross-section. \\
The asymptotic conical structure can be seen as a persistent interaction after a transient local interaction with curvature. This implies that the spacetime's asymptotic configuration retains information about the vortex matter content. In \cite{silva2021scattering}, we showed that it is needed to insert this long-term interaction inside the unscattered wave in the asymptotic field ansatz
\begin{equation}
\phi_\text{ansatz} = f(\varphi) \frac{e^{ip' w}}{\sqrt{r}} + \sum_{m=-\infty}^{\infty} A_m i^{|m|} J_{|m'|}(p'w) e^{im\varphi},
\end{equation}
with $A_m$ to be adjusted with the asymptotic field solution at infinity, eq. \eqref{scalar_asymp_solution}.
\\
Following the procedure in \cite{silva2021scattering}, the resulting scattering amplitude is given by

\begin{equation}
    f(\varphi) = \frac{1}{\sqrt{2\pi i p^\prime}}
    \sum_{m=-\infty}^\infty C_m e^{-i d_m} \left[e^{2id_m(p)} - 1 \right]e^{i (m\varphi - |m| \delta\varphi)} = \sum_{m=-\infty}^\infty f_m(\varphi),
    \label{scalar_SA}
\end{equation}
where $\delta \varphi = \frac{\pi}{2} \left(\frac{1}{b} - 1 \right)$, and $m$ in the coupling of $\delta \varphi$ replaced by $|m|$ as mentioned before. This leads to the total cross-section in the form
\begin{equation}
\frac{d \sigma}{d \varphi^\prime} = \frac{p^\prime}{p} |f(\varphi)|^2 \Rightarrow \sigma = \frac{4}{p} \sum_{m=-\infty}^{\infty} |C_m|^2 \sin^2(d_m). 
\label{sigma_scalar}
\end{equation}
Equation (\ref{sigma_scalar}) indicates that to determine the total cross-section, we require not only the phase $d_m$, which is typically sufficient in the standard partial wave approach, but also information regarding the amplitude of the scattered field.\\

\begin{figure}[H]
\includegraphics[width=1.0\textwidth]{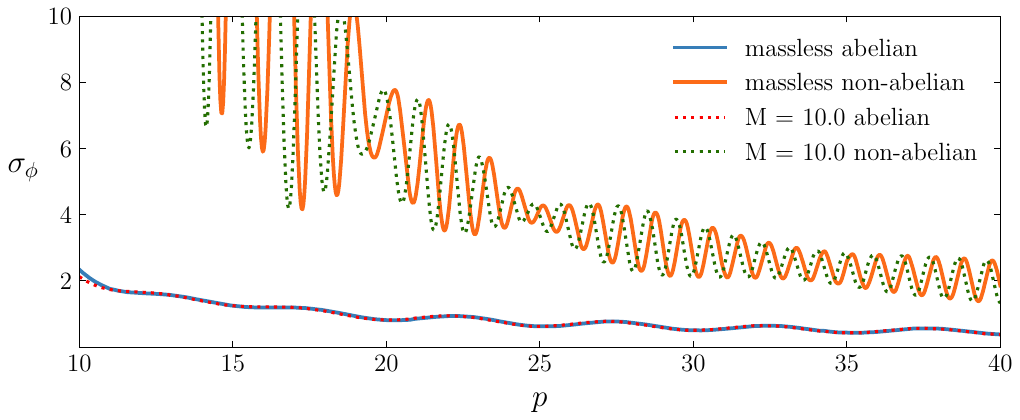}
\caption{Scattering cross-sections in abelian and non-abelian strings, choosing M = 0.0 and M = 10.0. Notice the oscillations present in both scenarios, though more evident in the non-abelian case.}
\label{sigma_scalar_abvsna}
\end{figure}

In order to solve the field equation (\ref{scalar_eom2}), we employed the 8th-order Runge-Kutta method already implemented in Python, via the \verb|scipy.integrate| module \verb|solve_ivp|, and then extracted the factors $C_m$, $d_m$ from the asymptotic solution. The numerical algorithm is given in Appendix 1. Using the formula (\ref{sigma_scalar}), we obtain the total scattering cross-section for the abelian and non-abelian strings depicted in Fig. \ref{sigma_scalar_abvsna}. In this figure, we present the results for both massless and massive scalar fields, choosing a specific value for the mass as an example.

We can see that the cross-section diverges as the incident momentum $p$ approaches zero and tends to zero as $p \rightarrow \infty $, which matches our intuition. Additionally, the presence of mass introduces a ``momentum delay'', shifting the plot backward in momentum space. However, an unusual oscillatory pattern emerges in the total cross-section in both scenarios. These oscillations are tied to the metric parameters in \eqref{metric_coniclimit}. Our hypothesis is that the physical mechanism behind this effect is diffraction, i.e., the scattered wave experiences diffraction when it encounters the conical, rigid, flat, asymptotic structure. 

\begin{figure}[H]
\includegraphics[width=0.9\textwidth]{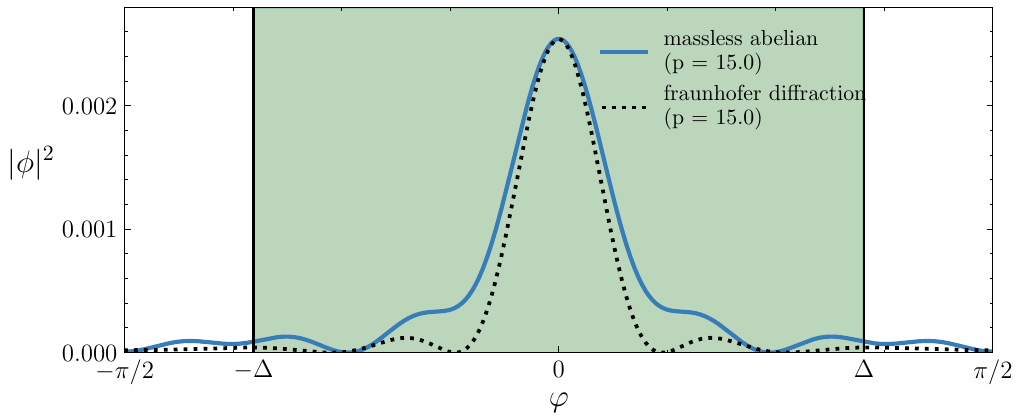}
\caption{Comparison of the field profiles of a scattered massless scalar plane-wave in the abelian case and a single-slit light diffraction in the Fraunhofer limit. The incident momentum is the same, $p = 15$. $\Delta = \pi (1 - b)$ corresponds to the abelian case, and the effective infinity is at $r = 600$.}
\label{scalar_field_inf_diff}
\end{figure}

To justify our hypothesis, let us consider single-slit light diffraction. When an electromagnetic plane-wave of frequency $\omega$, propagating in the $\hat{x}$-direction, hits an aperture of size $a_p$, placed at the origin of the x-y plane, the resulting electric field configuration far from the aperture is expressed as
\begin{equation}
E = 2 \frac{e^{i \omega t}}{\sqrt{r}} a_p \frac{ \sin(\frac{1}{2} \omega a_p \sin\varphi)}{\omega a_p \sin\varphi} \quad .
\label{fraunhofer_field}
\end{equation}
Now we can compare \eqref{fraunhofer_field} with the scattered massless scalar field profile far from the vortex. The abelian case serves as the most suitable comparison due to the minimal alteration of the initial plane wave by curvature. The result is shown in Fig. \ref{scalar_field_inf_diff}.

In \cite{fernandez2017emergence}, the authors showed that the conical topology could cause diffraction patterns when a scalar plane wave interacts with an ideal cosmic string. Figure \ref{scalar_field_inf_diff} illustrates a similar diffraction-like pattern observed when a scalar wave interacts with an extended vortex. The study in \cite{fernandez2017emergence} highlights that the angular region $[-\Delta, +\Delta]$, where $\Delta =\pi (1 - b)$, is where diffraction effects are important and geometrical optics fails. This finding holds reasonably well in the non-abelian case, with a larger angular range $\Delta$.

It is worth mentioning that the field angular profile should not always be very similar or well fitted to a Fraunhofer diffraction profile, as the field also interacts with the gravitational potential before reaching the asymptotic geometry (\ref{metric_coniclimit}). In the abelian case, the Fraunhofer diffraction model fits well, indicating that transient interactions with local curvature have minimal impact on the asymptotic configuration. This is evident from the peak of $|\phi|^2$ at $\varphi = 0$ in Fig. \ref{scalar_field_inf_diff}, the same direction of the incident wave, suggesting that the transient curvature is not strong enough to alter the particle flux significantly.

\begin{figure}[H]
\includegraphics[width=1.0\textwidth]{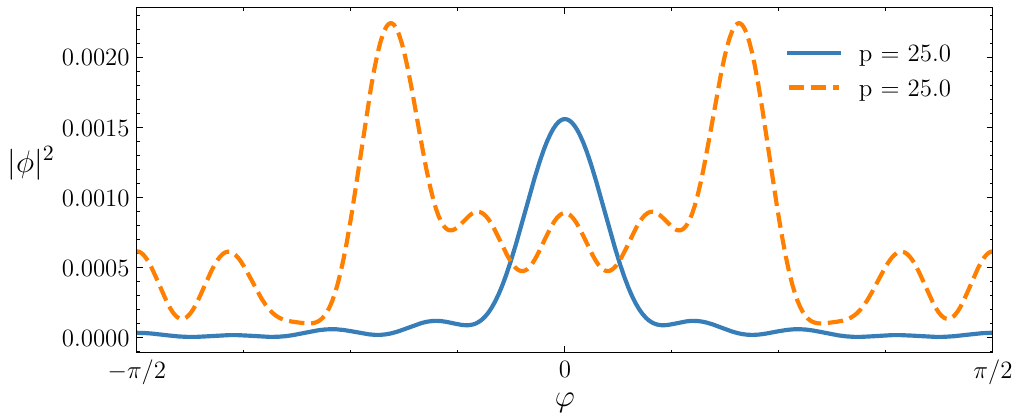}
\caption{Angular profile of the modulus squared of the field at infinity, scattered from abelian (solid blue curve) and non-abelian (dashed orange curve) vortices. }
\label{field_inf_both}
\end{figure}

In Figure~\ref{field_inf_both}, we present the field profiles for both the abelian and non-abelian cases. In the non-abelian case, we observe that the peak of $ |\phi|^2 $ is not centered; instead, it splits into two symmetric peaks around $ \varphi = 0$. This behavior can still be interpreted as a diffraction effect, arising from a narrower effective aperture in the non-abelian case and a more localized flux in the $ y $-direction. The uncertainty principle then leads to the central peak splitting into two. This interpretation is supported by the fact that the two peaks move closer to $ \varphi = 0 $ as the incident momentum increases.

\begin{figure}[H]
\centering
\includegraphics[width=0.8\textwidth]{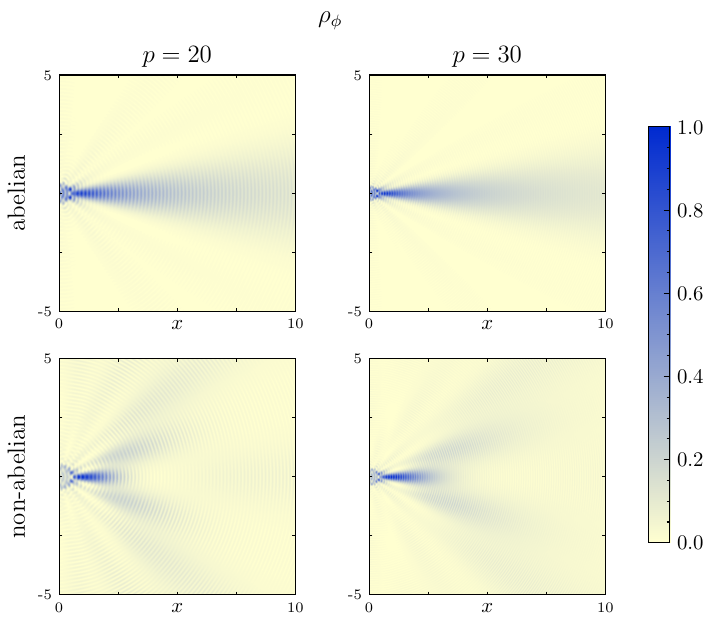}
\caption{In the abelian scenario most of the particles are concentrated around $\varphi = 0$, whilst in the non-abelian scenario the initial flux around $\varphi = 0$ is dispersed to other directions, consistent with the diffraction picture.}
\label{scalar_heatmap_comparison}
\end{figure}

This effect is already evident in the Fresnel regime, i.e., at short distances from the core, as shown in Figure~\ref{scalar_heatmap_comparison}, where we plot the normalized profile of $ \rho_{\phi}=|\phi|^2 $ for $ x \leq 10 $ and $|y| \leq 5$. The normalization is performed with respect to the highest value of $\rho_{\phi}$ found in the plotted domain.

\section{Fermion Field}
\label{fermion}
In order to analyze the scattering of a massive spinor field, we first need to solve the Dirac equation coupled with gravity. We then propose an ansatz that typically encodes the symmetry of our problem, subsequently reducing the Dirac equation to a set of coupled ordinary differential equations. These equations are then solved numerically. Once we have this solution, we can plot the angular profile and extract all the necessary information to construct the cross-section, just as we did with the scalar field.

The Dirac equation in curved spacetime is given by 
\begin{equation}
(\gamma^A e^\mu \, _A \partial_ \mu - \gamma^A \Gamma_A + iM_f)\Psi = 0,
\label{Dirac-eq}
\end{equation}
where $\Gamma_A$ and $M_f$ are the spin connection and the fermion mass, respectively. The Greek indices are lowered and raised by the spacetime metric $g_{\mu \nu}=\mathrm{diag}(N^2,-1,-L^2,-N^2)$ and the Latin indices by $\eta_{AB}=\mathrm{diag}(1,-1,-1,-1)$. 
Let us start with the set of tetrad one forms $\{ \omega^A = e^A \, _\mu dx^\mu \}$ 
\begin{align}
\begin{split}
&\omega^0 = Ndt,\\
&\omega^1 = \cos{\phi} d\rho + \sin{\phi}L d\phi,\\
&\omega^2 = -\sin{\phi} d\rho + \cos{\phi}L d\phi,\\
&\omega^3 = N dz,
\end{split}
\end{align}
or equivalently $\{ dx^\mu = e^\mu \, _A \omega^A \}$ 
\begin{align}
\begin{split}
&dt = \omega^0/N,\\
&d\rho = \omega^1 \cos{\phi} - \omega^2 sin{\phi},\\
&d\phi = (1/L)(\omega^1 \sin{\phi} + \omega^2 \cos{\phi}),\\
&dz = \omega^3 /N \, ,
\end{split}
\end{align}
giving rise to
\begin{align}
ds^{2}&=(\omega^0)^{2}-(\omega^1)^{2}-(\omega^2)^{2}-(\omega^3)^{2} \nonumber\\
&= \eta_{AB} \omega^A \omega^B. \label{ds1}
\end{align}
As a result one can read the tetrad matrices as
\begin{align}\label{tetrad}
e^{\mu} \, _A &= 
\begin{pmatrix}
1/N & 0 & 0 & 0\\
0 & \cos{\phi} & -\sin{\phi} & 0\\
0 & (1/L)\sin{\phi} & (1/L)\cos{\phi} & 0\\
0 & 0 & 0 & 1/N\\ 
\end{pmatrix},\\
e^A \, _{\mu} &= 
\begin{pmatrix}
N & 0 & 0 & 0\\
0 & \cos{\phi} & L\sin{\phi}  & 0\\
0 & -\sin{\phi} & L\cos{\phi} & 0\\
0 & 0 & 0 & N\\
\end{pmatrix}.
\label{tetrad2}
\end{align}
The spin connection is given by
\begin{equation}
\Gamma_C = -\frac{1}{4} \gamma_{ABC} \gamma^A \gamma^B,
\end{equation}
where 
\begin{equation}
\gamma^A = \frac{1}{2} (C_{ABC} + C_{BAC} - C_{CAB}) \, .
\end{equation}
To obtain $C_{ABC}$ coefficients one can use Cartan's first structure equation 
\begin{equation}
d\omega^A = \frac{1}{2}C^A \, _{BC} \, \omega^B \wedge \omega^C.
\end{equation}
The exterior derivatives of the tetrad one forms are as follows
\begin{align*}
\begin{split}
d\omega^0 &= \frac{N'}{N} \Big{(} \cos{\phi} \, \omega^1 \wedge \omega^2 - \sin{\phi} \, \omega^2 \wedge \omega^0 \Big{)},\\
d\omega^1 &= \frac{1}{L} \Big{(} L' +  1\Big{)} \, \sin{\phi} \, \omega^1 \wedge \omega^2,\\
d\omega^1 &= \frac{1}{L} \Big{(} L' + 1\Big{)} \, \cos{\phi} \, \omega^1 \wedge \omega^2,\\
d\omega^3 &= \frac{N'}{N} \Big{(} \cos{\phi} \, \omega^1 \wedge \omega^3 - \sin{\phi} \, \omega^2 \wedge \omega^3 \Big{)}.
\end{split}
\end{align*}
Finally, we choose the following representation for the gamma matrices $\gamma^A$ 
\begin{align}
\gamma^0 = 
\begin{pmatrix}
\boldsymbol{1} & 0\\
0 & -\boldsymbol{1}
\end{pmatrix}
\quad \quad \textrm{and} \quad \quad
\gamma^i = 
\begin{pmatrix}
0 & \boldsymbol{\sigma^i}\\
-\boldsymbol{\sigma^i} & 0
\end{pmatrix}.
\end{align}
Putting all the above information together gives the spin connection
\begin{align}
\begin{split}
\Gamma_0 &= - \frac{N'}{2N}
\begin{pmatrix}
 0 & \chi \\
\chi & 0 \\
\end{pmatrix},\\
\Gamma_1 &= \frac{i}{2L}(L' + 1)\sin{\phi}
\begin{pmatrix}
\sigma^3  & 0\\
 0 & \sigma^3\\
\end{pmatrix},\\
\Gamma_2 &= \frac{i}{2L}(L' + 1)\cos{\phi}
\begin{pmatrix}
\sigma^3  & 0\\
 0 & \sigma^3\\
\end{pmatrix},\\
\Gamma_3 &= \frac{N'}{2N}
\begin{pmatrix}
\eta & 0\\
0 & \eta \\
\end{pmatrix},\\
\end{split}
\end{align}
with
\begin{align}
\begin{split}
\eta= 
\begin{pmatrix}
 0 & -e^{i \phi}\\
  e^{-i\phi} & 0\\
\end{pmatrix}
\end{split}
\quad \quad \textrm{and} \quad \quad
\chi= 
\begin{pmatrix}
 0 & e^{i \phi}\\
  e^{-i \phi} & 0\\
\end{pmatrix}.
\end{align}
Now we need to insert the spin connection together with the tetrad matrix (\ref{tetrad}) into the Dirac equation (\ref{Dirac-eq}). We employ the following ansatz
\begin{equation}
\Psi(t,\rho,\phi,z)= \sum_{j = - \infty}^{\infty} \Psi_j(t, \rho, \varphi, z)=e^{-iEt}e^{ikz}\sum_{j = -\infty}^{\infty}a_j\psi_j(\rho,\varphi)e^{ij\varphi} ,
\end{equation}
with $j = \pm 1/2, \pm 3/2, \pm 5/2 ...$ and

\begin{equation}
\psi_j(\rho, \varphi) = 
\begin{pmatrix}
\psi^{(0)}(\rho) e^{-i \varphi/2} \\
\psi^{(1)}(\rho) e^{+i \varphi/2} \\
\psi^{(2)}(\rho) e^{-i \varphi/2} \\
\psi^{(3)}(\rho) e^{+i \varphi/2} 
\end{pmatrix}
,
\label{fermion_ansatz2}
\end{equation}
where we have dropped the index $j$ from the components $\psi_j(\rho) $ of the spinor $\psi_j(\rho,\varphi)$. 
The angular dependence in (\ref{fermion_ansatz2}) is such that it matches the Minkowski solution in the limit $N\to 1$ and $L\to \rho$. Besides that, we consider the dimensionless parameters $\sqrt{\lambda_1}\, \eta_1 \, \rho \equiv r$, $E/\sqrt{\lambda_1}\, \eta_1 \,\equiv \tilde E$, $k/\sqrt{\lambda_1}\, \eta_1 \,\equiv\tilde k$ and $M_f/\sqrt{\lambda_1}\, \eta_1 \, \equiv\tilde M_f$ which leads to the following differential equations for the spinor components
\begin{equation}
\begin{pmatrix}
(i/N) \big{(} -\tilde E \psi^{(0)} + \tilde k \psi^{(2)} \big{)} + i\tilde M_f \psi^{(0)} +  [ \partial_r + (j+1/2)/L + \xi(r) ]\psi^{(3)}\\
(i/N) \big{(} -\tilde E \psi^{(1)} - \tilde k \psi^{(3)} \big{)} + i\tilde M_f \psi^{(1)} + [ \partial_r - (j-1/2)/L+ \xi(r)]\psi^{(2)} \\
(i/N) \big{(} \tilde E\psi^{(2)} - \tilde k \psi^{(0)} \big{)} + i\tilde M_f \psi^{(2)} - [ \partial_r + (j+1/2)/L + \xi(r)]\psi^{(1)} \\
(i/N) \big{(} \tilde E \psi^{(3)} + \tilde k \psi^{(1)} \big{)} + i\tilde M_f \psi^{(3)} - [ \partial_r - (j - 1/2)/L + \xi(r)]\psi^{(0)} \\
\end{pmatrix}
= \boldsymbol{0},
\label{fermion_eom1}
\end{equation}
where we have defined $\xi(r)\equiv \big{[} N'/N + (1/2L)(L'-1) \big{]}$. This parameter vanishes in Minkowski space but not in the conical one. All the above $\psi^{(i)}$ are complex functions and only dependent on $r$. From now on, we will drop the tilde.

In order to employ the appropriate partial-wave approach, it is crucial to understand the solution of (\ref{fermion_eom1}) in the limits $r \rightarrow 0$ and $r \rightarrow \infty$ which is essential for formulating a consistent asymptotic ansatz for each component of the solution.
In \cite{mohammadi2015finite} it was shown that the positive-energy solution of \eqref{fermion_eom1} when $r \rightarrow 0$ is given by
\begin{equation}
\Psi_j(t, r \rightarrow 0, \varphi, z) = a_j e^{-iEt} e^{ikz} e^{ij\varphi}
\begin{pmatrix}
J_{\beta_j}(p r)e^{-i\varphi/2} \\
s J_{\beta_j + \epsilon_j}(p r)e^{i\varphi/2} \\
\frac{k - s i \epsilon_jp}{E + M} J_{\beta_j}(p r)e^{-i\varphi/2} \\
-s \frac{k - s i \epsilon_jp}{E + M} J_{\beta_j + \epsilon_j}(p r)e^{i\varphi/2}\\
\end{pmatrix},
\end{equation}
where $j = \pm 1/2, \pm 3/2, ...$, $\epsilon_j = \text{sgn(j)}$, $s=\pm 1$ the helicity, $\beta_j = |j| - \epsilon_j/2$ and $p^2 = E^2 - k^2 - M_f^2$ as previously defined. The proportionality constant $a_j$ is determined by the chosen initial condition, following a similar approach to the scalar field. Here we set $a_j = i^{\left| j - \frac{1}{2} \right|}$ and $k = 0$ such that, when we average over the helicity, the spinor field $\Psi$ is a plane-wave in the $\hat{x}$-direction near the origin 
\begin{equation}
\Psi(t, r \to 0, \varphi, z) = e^{-i E t}
\begin{pmatrix}
1 \\
0 \\
0\\
\frac{p}{E + M_f} \\
\end{pmatrix}
e^{i p r \cos\varphi} .
\label{fermion_initcond}
\end{equation}
Upon examining \eqref{fermion_eom1}, we note that with the initial condition \eqref{fermion_initcond}, the components $\psi^{(1)}$ and $\psi^{(2)}$ vanish identically, causing the second and third rows in \eqref{fermion_eom1} to also vanish. By defining the following operator
\begin{equation}
\nabla_{\pm} = \partial_r \pm \frac{j + 1/2}{L(r)} + \xi(r),
\end{equation}
 the resulting equations of motion can be represented in a rather symmetric form
\begin{equation}
\begin{aligned}
\nabla_+ \psi^{(3)} &= i \left(\frac{E}{N} - M_f \right) \psi^{(0)}, \\
\nabla_- \psi^{(0)} &= i \left(\frac{E}{N} + M_f \right) \psi^{(3)}.
\end{aligned}
\end{equation}
To determine the form of the solution far from the vortex, we observe that in the asymptotic limit \eqref{metric_coniclimit}, the equations of motion \eqref{fermion_eom1} take on a form similar to those in Minkowski space, although with $r, j, E$ replaced by $w = r + c/b, j' = j/b, E' = E/a$. This substitution leads to the following asymptotic solution 
\begin{equation}
\psi_j(r \rightarrow \infty, \varphi) = C_j \sqrt{\frac{2}{\pi p^\prime r}}
\begin{pmatrix}

\cos \left[ p^\prime w - \frac{\pi}{2}\left(\beta_{j^\prime} + \frac{1}{2} \right) + d_j^0 \right] e^{-i \varphi/2}\\

0 \\

0 \\

\frac{i\epsilon_{j^\prime} p^\prime}{E^\prime + M} \cos \left[ p^\prime w - \frac{\pi}{2}\left(\beta_{j^\prime}+\epsilon_{j^\prime} + \frac{1}{2} \right) + d_j^3 \right] e^{+i \varphi/2}\\
\end{pmatrix}
,
\label{fermion_asymp_solution}
\end{equation}
where ${p^\prime}^2 = {E^{\prime}}^2 - M_f^2$, just as in the scalar scenario with $k=0$. $C_j$ and $d_j^i$ are determined numerically by considering the interaction of the field with the asymptotic structure.
Now, we can separate the problem into the scattering of two spinor components satisfying the initial condition \eqref{fermion_initcond}. The asymptotic ansatz takes the general form
\begin{equation}
\psi_{\text{ansatz}}^i = f^i(\varphi) 
\begin{pmatrix}
1 \\
0 \\
0 \\
\frac{p'}{E' + M_f}
\end{pmatrix}
\frac{e^{ip'w}}{\sqrt{r}}
 + (\psi_{mpv})^i .
\label{fermion_ansatz}
\end{equation} 
The last term represents the corresponding modified plane waves, with $i = 0, 1, 2, 3$ labeling the components of the spinor field. We have explicitly presented the components of the spinorial cylindrical wave. For consistency, we define the modified plane-wave ansatz as follows
\begin{equation}
\psi_{mpv}= e^{-iEt} \sum_{j = -\infty}^{\infty}
\begin{pmatrix}
A_j^0 i^{\left|j - \frac{1}{2} \right|} J_{\beta_{j^\prime}}(p^\prime w)e^{i \left(j - \frac{1}{2} \right)\varphi} \\
0 \\
0 \\
A_j^3 i^{\left|j - \frac{1}{2} \right|} J_{\beta_{j^\prime}+\epsilon_{j^\prime}}(p^\prime w)e^{i\left(j + \frac{1}{2} \right) \varphi} \\
\end{pmatrix}
.
\label{fermion_mpw_ansatz} 
\end{equation}
The constants $A_j^i$ are to be determined from the asymptotic solution. With this, we can compute the scattering amplitudes $f^{i}(\varphi)$. Equating $\psi^{(0)}_{\text{ansatz}}$ with $\psi^{(0)} (r \to \infty)$ and expanding the cosines in complex exponentials, we obtain $A^0_j = C_j e^{-i d^0_j}$, leading to
\begin{equation}
f^{(0)}(\varphi) = \frac{1}{\sqrt{2 \pi i p'} } \sum_{j=-\infty}^{\infty} i^{ \left| j - \frac{1}{2} \right|} C_j e^{- i d_j^0} \left( e^{2 i d_j^0} - 1 \right) e^{-i \frac{\pi}{2} \beta_{j'}} 
e^{i \left(j - \frac{1}{2} \right) \varphi}.
\end{equation}
Replacing $\beta_j = \left|j - \frac{1}{2} \right|$, the expression reduces to
\begin{equation}
f^{(0)}(\varphi) = \frac{e^{-i \varphi/2}}{\sqrt{2 \pi i p'} } \sum_{j=-\infty}^{\infty} C_j e^{- i d_j^0} \left( e^{2 i d_j^0} - 1 \right) e^{i \left(j\varphi -|j| \delta\varphi \right)}.
\label{fermion_SA0}
\end{equation}
Therefore, the total cross-section associated with the 0-th component becomes
\begin{equation}
\sigma^{(0)} = \frac{4}{p} \sum_{j = -\infty}^{\infty} |C_j|^2 \sin^2(d_j^0).
\label{sigma_fermion_0}
\end{equation}

Now, for the third component, equating $\psi^{(3)}_{\text{ansatz}} = \psi^{(3)} (r \to \infty)$,  and substituting $\beta_j + \epsilon_j = \left| j + \frac{1}{2} \right| $ we obtain
\begin{equation}
\begin{aligned}
f^{(3)}(\varphi) &= \frac{e^{i \varphi/2}}{\sqrt{2 \pi i p'}} \sum_{}^{}
C_j e^{-i d_j^3} \left( e^{2 i d_j^3} - 1 \right)
e^{i \left(j \varphi - |j| \delta \varphi \right)}, \\
\sigma^{(3)} &= \frac{4}{p} \sum_{j = -\infty}^{\infty} |C_j|^2 \sin^2(d_j^3).
\end{aligned}
\end{equation}

As the initial condition is already helicity-averaged, the total cross-section for the spinor field is the sum of the individual cross-sections of each component. Thus, the total cross-section for the spinor field is 
\begin{equation}
\sigma_{\Psi} = \frac{4}{p} \left[  \sum_{j = -\infty}^{\infty} |C_j|^2 \sin^2(d_j^0) + \sum_{j = -\infty}^{\infty} |C_j|^2 \sin^2(d_j^3) \right],
\end{equation}
which is equivalent to the sum of the total cross-sections of two scalar fields with the same asymptotic amplitude but different phase shifts.

In order to obtain the fermionic total cross-section in the spacetimes depicted in Fig. \ref{metric}, we need to solve the equations of motion \eqref{fermion_eom1} numerically and extract the amplitudes and phase-shifts from the asymptotical solution, similarly to what we did in the scalar case. We solved the equations of motion \eqref{fermion_eom1} using the 8th-order Runge-Kutta method via the \verb|scipy.integrate| module \verb|solve_ivp|. The algorithm used to extract the asymptotic parameters is the same as the one in the scalar scenario, and it is detailed in Appendix 1.

Figure~\ref{fermion_cross} shows the total fermionic cross-section in both the abelian and non-abelian cases. Oscillatory behavior is evident, similar to the pattern observed for the scalar field. While the qualitative structure of Fig. \ref{fermion_cross} resembles that of Fig. \ref{sigma_scalar_abvsna} for the scalar field scattering, it is important to emphasize that the scales differ considerably. This indicates that, under the same set of parameters, the spinor field exhibits a broader scattering area compared to the scalar field.

\begin{figure}[H]
\includegraphics[width=1.0\textwidth]{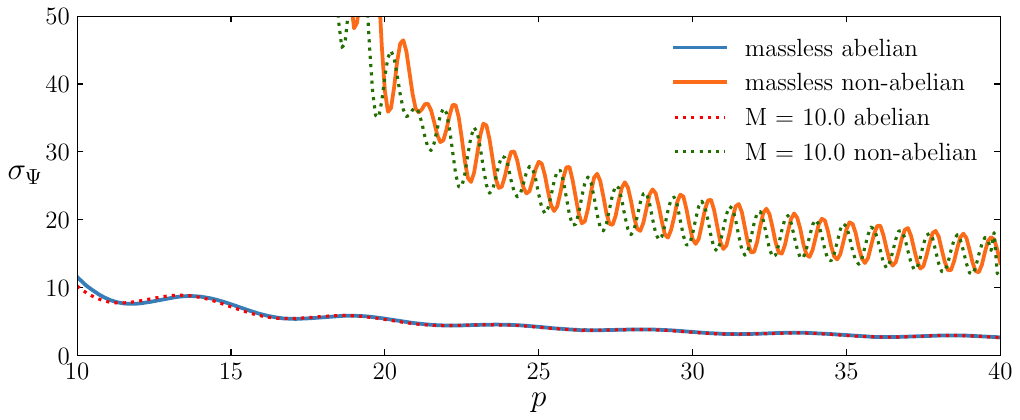}
\caption{Fermionic total cross-section for both abelian and non-abelian cases. Again, we see oscillations in both cases, and the overall amplitude in the non-abelian scenario is more significant than in the abelian case.}
\label{fermion_cross}
\end{figure}

\begin{figure}[H]
\includegraphics[width=1.0\textwidth]{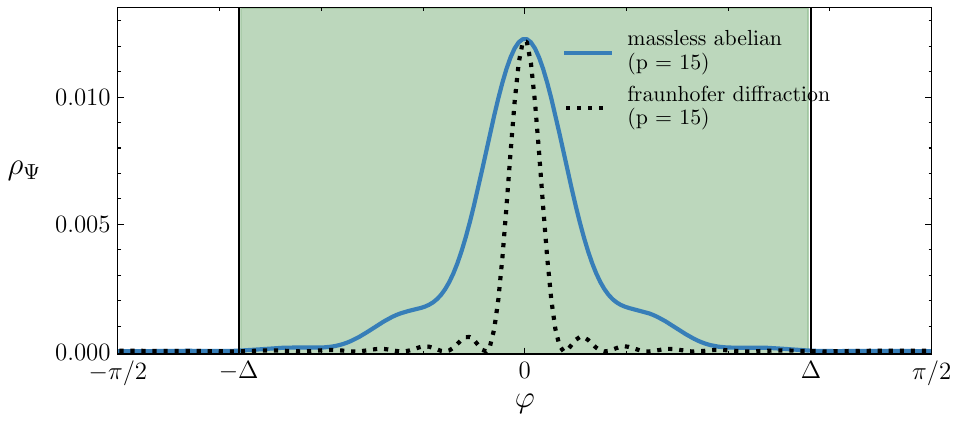}
\caption{Diffraction pattern for the spinor field in the abelian scenario. Here, we show the time component of the 4-current, $J^0 = \rho_{\Psi} =\bar{\Psi} \gamma^0 \Psi$, at $r=600$.}
\label{fermion_field_inf_fraunhofer}
\end{figure}

Regarding the asymptotic angular profile, we observe that the spinor field also undergoes diffraction, as depicted in Fig. \ref{fermion_field_inf_fraunhofer} showing the angular profile of the 0-th component of the 4-current, $J^0 = \rho_{\Psi} = \bar{\Psi} \gamma^0 \Psi = \Psi^\dagger \Psi$. However, the Fraunhofer diffraction profile does not match as closely as it did for the scalar field. 

\begin{figure}[H]
\includegraphics[width=1.0\textwidth]{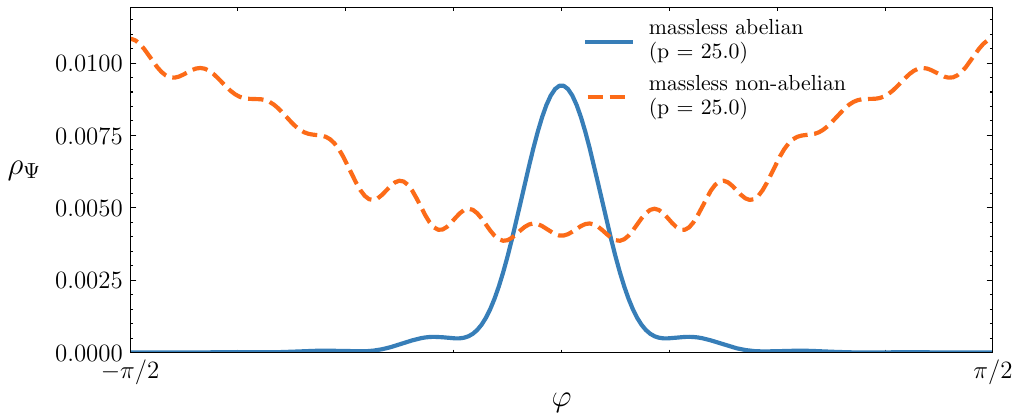}
\caption{The fermionic diffraction profile is similar to the scalar profile.}
\label{fermion_field_inf}
\end{figure}

Comparing the massless fermionic angular profiles interacting with the abelian and non-abelian vortex spacetimes, shown in Fig. \ref{fermion_field_inf}, we observe notable similarities with those of the scalar field. In the abelian case, the peak of $\rho_{\Psi}$ also aligns with $\varphi = 0$, and in the non-abelian case, we observe that the central peak is split around $\varphi = 0$. Similar to the scalar case, in the non-abelian scenario, the peaks of $\rho_{\Psi}$ progressively converge to $\varphi = 0$ as the incident momentum increases. This can be interpreted similarly to before; the uncertainty principle leads to splitting the central peak due to increased localization, or equivalently, narrower effective aperture, in the $y$-direction.

\begin{figure}[H]
\centering
\includegraphics[width=0.8\textwidth]{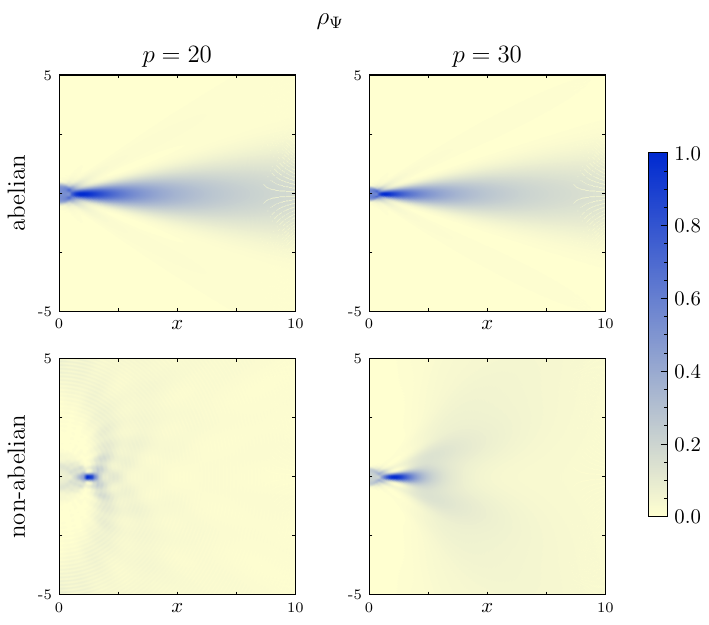}
\caption{Fermionic particles, while still concentrating in the $\varphi = 0$, are more scattered than scalar particles even in the abelian scenario. Also, the characteristic peak-splitting observed in the scalar field profile is also present here.}
\label{fermion_heatmap_comparison}
\end{figure}

Moreover, the dispersion around $\varphi = 0$ is already visible in the Fresnel regime, as illustrated in Fig.~\ref{fermion_heatmap_comparison}, where the normalized $\rho_\Psi$ is shown for $x \leq 10$ and $|y| \leq 5$. Again, normalization is performed with respect to the maximum value of $\rho_\Psi$ within the displayed grid. 

\section{Conclusion}
\label{conclusion}

In this work, we examined the scattering of scalar and spinor fields in the background of a gravitating cosmic string spacetime originating from a non-abelian vortex. The spacetime is characterized by the metric \eqref{metric_ansatz} with the coefficients specified in Fig. \ref{metric}. We also compared the results with the corresponding abelian cosmic string, taking the appropriate limit within the same model.

We first focused on scalar field scattering, employing the modified partial-wave method developed in Ref.\cite{silva2021scattering}. We computed both the total cross-section and the asymptotic angular distribution of the field. The total cross-section presents distinctive oscillations, which we attribute to diffraction effects caused by the asymptotic conical geometry of the spacetime \eqref{metric_coniclimit}. We have shown that the field angular profile follows a pattern similar to a Fraunhofer diffraction, as depicted in Fig. \ref{scalar_field_inf_diff}. 
In the second part, we extended the analysis to spinor fields. Initially, we developed the partial-wave theory tailored for the spinor field, following a framework similar to that of the scalar case. The total cross-section and angular profiles for the fermionic field also reveal oscillatory features. However, the overall magnitude of the fermionic cross-section is notably larger, indicating that fermionic particles experience stronger scattering in the same background geometry. Although a diffraction pattern is still evident in the fermionic case, it deviates more from the ideal Fraunhofer form compared to the scalar field.

For both scalar and fermion fields, the angular density peaks at $\varphi = 0$ in the abelian scenario, whereas in the non-abelian case, the central peak is split around $\varphi = 0$.
This splitting arises due to the larger deficit angle associated with the non-abelian string, which effectively narrows the aperture and increases spatial localization.
As the field becomes more localized in space, the uncertainty in momentum increases, causing the field to spread in other directions, becoming less localized around the center $\varphi = 0$. We also noticed that the diffraction picture is already present at the Fresnel regime, i.e., at small distances from the core, agreeing with Fraunhofer's asymptotic diffraction result. We observed that the spinor field consistently exhibits a broader angular spread than the scalar field for the same spacetime background, as seen in Figs. \ref{scalar_heatmap_comparison} and \ref{fermion_heatmap_comparison}.

Finally, our results offer a pathway to a detailed account of matter field scattering in realistic cosmic string spacetimes, highlighting key differences between scalar and fermionic interactions. These findings may contribute to a deeper understanding of potential observational signatures associated with cosmic strings and their interaction with surrounding matter.

\appendix
\section{Numerical algorithm to find $C_m$ and $d_m$}

In all scenarios shown in this work, the asymptotic radial solution is in the form

\begin{equation}
y = \frac{C}{\sqrt{r}} \cos(pr + d).
\end{equation}
In order to extract the constants $C$ and $d$, we first remove the damping factor $\sqrt{r}$

\begin{equation}
y' = \sqrt{r}y = C \cos(pr + d).
\end{equation}
Now we notice that the maximum value of $y'$ is $|C|$, that is

\begin{equation}
\max(y') = \max( C \cos(pr + d) ) = |C|.
\end{equation}
The amplitude $|C|$ and the phase $d$ give sufficient information to compute the total cross-section.

Given the solution $y'$ we calculate $d$ in two steps. In the first step we find the phase by finding the distances between the peaks of $f(r) = y'$ and a function $g(r, \delta) = |C| \cos(pr + \delta)$, with $\delta = 0$. As a measure of error we calculate the correlation between $f(r)$ and $g(r)$ defined as

\begin{equation}
\text{corr}(f,g) \equiv \frac{\int{f(x)g(x) dx}}{\sqrt{\int{f(x)^2 dx}} \sqrt{\int{g(x)^2} dx}} = \cos(d - \delta),
\end{equation}
where the integrals are performed in a cycle of $f(r)$, i.e., $[r_0$, $r_0 + 2\pi/p] $. If the correlation is below a threshold, $\text{corr}_0$, we go to the second step, in which we minimize the residue function $\text{res}(\delta) = |\text{corr}(f(r), g(r, \delta) )- 1|$ with respect to $\delta$. This is easily done using the \verb|scipy.optimize| module \verb|minimize_scalar|. We typically found values of the residue function to be around $10^{-5}$.

We extracted the parameters, in all scenarios, at around $r_0 = 600$ with $\text{corr}_0 = 0.96$.

\section*{Acknowledgements}
AM and MS thank Antônio de Pádua Santos and Eugênio R Bezerra de Mello for providing the data used in \cite{de2015gravitating}.
AM acknowledges financial support from Conselho Nacional de Desenvolvimento Científico e Tecnológico (CNPq), Grant no. 306295/2023-7. AM and MS acknowledge financial support from Coordenação de Aperfeiçoamento de Pessoal de Nível Superior (CAPES). Part of the simulations exhibited here were performed in the supercomputer SDumont of the Brazilian agency LNCC (Laboratório Nacional de Computação Científica). 

\printbibliography
\end{document}